\documentclass{article}

\PassOptionsToPackage{numbers}{natbib}

\usepackage[final]{neurips_2023_ml4ps}
 
\usepackage[utf8]{inputenc} 
\usepackage[T1]{fontenc}    
\usepackage{hyperref}       
\usepackage{url}            
\usepackage{booktabs}       
\usepackage{amsfonts}       
\usepackage{nicefrac}       
\usepackage{microtype}      
\usepackage{xcolor}         
\usepackage{graphicx}
\definecolor{seagreen}{rgb}{0.190, 0.525, 0.361}
\definecolor{midnightBlue}{rgb}{0.098, 0.098, 0.439}
\definecolor{darksalmon}{rgb}{0.914, 0.588, 0.478}

\title{\emph{Causa prima}: cosmology meets causal discovery for the first time}

\author{%
  \small{Mario Pasquato} \\
  \small{Département de Physique, Université de Montréal,}\\ \small{1375 Avenue Thérèse-Lavoie-Roux, Montreal, Canada}\\
  \small{Mila - Quebec Artificial Intelligence Institute,}\\ \small{6666 Rue Saint-Urbain, Montréal, Canada}\\
  \small{Ciela - Montreal Institute for Astrophysical Data Analysis}\\ \small{and Machine Learning, Montréal, Canada}\\
  \small{Dipartimento di Fisica e Astronomia, Università di Padova,}\\
  \small{Vicolo dell'Osservatorio 5, Padova, Italy}\\
  \small{INFN–Padova, Via Marzolo 8, Padova, Italy}\\
  \small{\texttt{mario.pasquato@umontreal.ca}}\\
     \vspace{-0.28in}
   \And
   \small{Zehao Jin} \\
   \small{Center for Astrophysics and Space Science (CASS),}\\ \small{New York University Abu Dhabi, PO Box 129188, Abu Dhabi, UAE}\\
   \vspace{-0.28in}
  \And
   \small{Pablo Lemos} \\
  \small{Département de Physique, Université de Montréal,}\\ \small{1375 Avenue Thérèse-Lavoie-Roux, Montreal, Canada}\\
  \small{Mila - Quebec Artificial Intelligence Institute,}\\ \small{6666 Rue Saint-Urbain, Montréal, Canada}\\
  \small{Ciela - Montreal Institute for Astrophysical Data Analysis}\\ \small{and Machine Learning, Montréal, Canada}\\
   \vspace{-0.28in}
   \And
   \small{Benjamin L. Davis} \\
   \small{Center for Astrophysics and Space Science (CASS),}\\ \small{New York University Abu Dhabi, PO Box 129188, Abu Dhabi, UAE} \\
      \vspace{-0.28in}
   \And
   \small{Andrea V. Macciò} \\
   \small{New York University Abu Dhabi, PO Box 129188, Abu Dhabi, United Arab Emirates }\\
   \small{Center for Astrophysics and Space Science (CASS),}\\ \small{New York University Abu Dhabi, PO Box 129188, Abu Dhabi, UAE} \\
   \small{Max-Planck-Institut für Astronomie, Königstuhl 17, 69117 Heidelberg, Germany} \\
 \vspace{-0.28in}
}

\begin{document}

\maketitle

\begin{abstract}
In astrophysics, experiments are impossible. We thus must rely exclusively on observational data. Other observational sciences increasingly leverage causal inference methods, but this is not yet the case in astrophysics. 
Here we attempt causal discovery for the first time to address an important open problem in astrophysics: the (co)evolution of supermassive black holes (SMBHs) and their host galaxies. We apply the Peter-Clark (PC) algorithm to a comprehensive catalog of galaxy properties to obtain a completed partially directed acyclic graph (CPDAG), representing a Markov equivalence class over directed acyclic graphs (DAGs). Central density and velocity dispersion are found to cause SMBH mass. We test the robustness of our analysis by random sub-sampling, recovering similar results. We also apply the Fast Causal Inference (FCI) algorithm to our dataset to relax the hypothesis of causal sufficiency, admitting unobserved confounds. Hierarchical SMBH assembly may provide a physical explanation for our findings.

\end{abstract}

\section{Introduction}
The puzzle of supermassive black hole (SMBH) and host galaxy coevolution is a long-standing open problem in astronomy \citep[][]{2005Natur.433..604D, 2013ARA&A..51..511K}
More recently, the detection of high-redshift quasars reignited the debate on the formation mechanisms of SMBHs in the early Universe \cite{2023arXiv230308918L}.
The main issue we face when trying to understand SMBH and host galaxy coevolution is the direction of causality: are galaxy properties driving SMBH assembly or is SMBH feedback shaping observed galaxy properties?
While it is likely that a feedback loop exists, with SMBH mass at a given time affecting galaxy properties at a later time and galaxy properties in turn affecting SMBH mass at an even later time, we only have data about a given galaxy and its SMBH at one point in time.

Given this situation, most of the research on this topic in astronomy has been based either on identifying observational correlations \citep[][]{1998AJ....115.2285M, 2000ApJ...539L..13G} or on developing models, based mainly on numerical simulations \citep[][]{2023MNRAS.525...12S}.
Here we break this pattern, introducing a method that to our knowledge has never before been applied in the field: we apply causal discovery to a state-of-the-art sample of galaxies.


\section{Data}
\begin{figure}
  \centering
  \includegraphics[width=\textwidth]{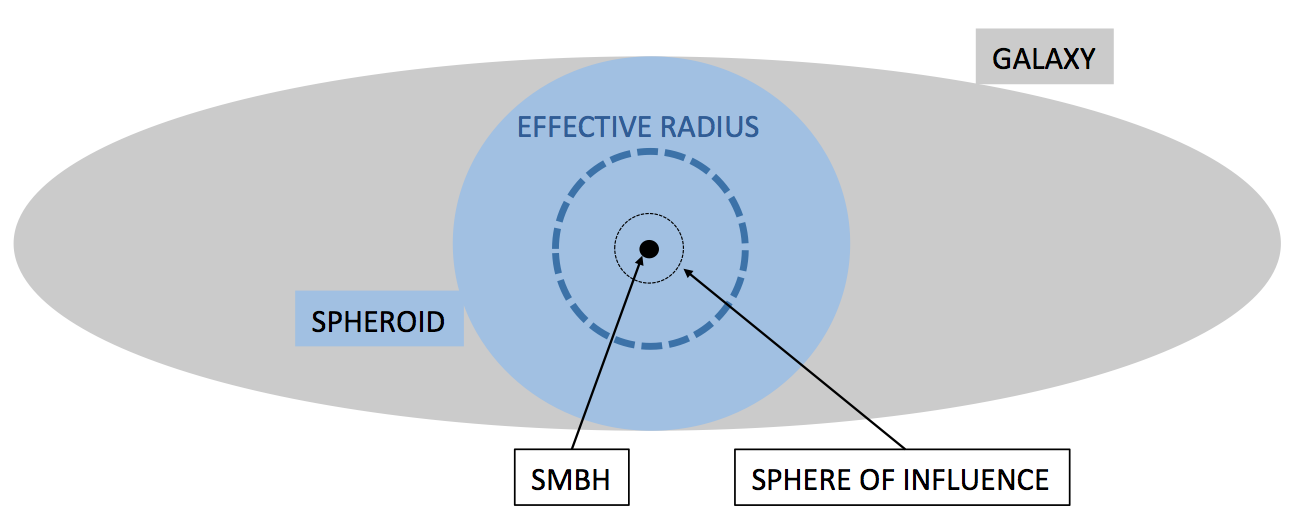}
  \caption{Early-type galaxy schematic, with a possible large-scale, rotationally-supported disk.
  The spheroid or bulge is the central, pressure (dispersion) supported part of the galaxy.
  The radius containing half its luminosity or effective radius gives a measure of its size.
  The SMBH is typically located in a central position.
  The sphere of influence of the SMBH corresponds to the region where stellar motions are dominated by the SMBH gravity.\label{galaxy}}
\end{figure}

\begin{figure}
  \centering
  \includegraphics[width=0.8\textwidth]{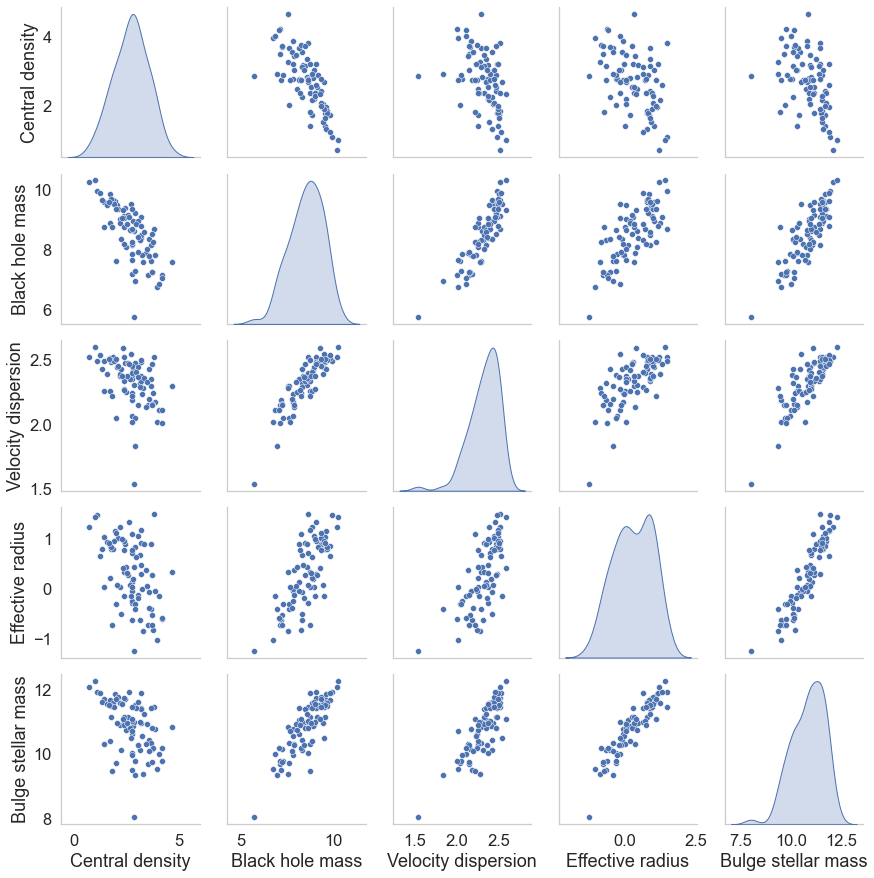}
  \caption{Pair plot of our data set of galaxy properties.\label{DataSet}}
\end{figure}

We consider a set of $83$ early-type galaxies with directly-measured SMBH masses compiled by \cite{2013ApJ...764..151G,2016ApJ...817...21S,2017MNRAS.471.2187D,2019ApJ...873...85D,2019ApJ...876..155S,2019ApJ...877...64D,2019ApJ...887...10S,2022ApJ...927...67S}.
The variables we consider are as follows: SMBH mass ($M_{BH}$), the stellar bulge (spheroid) mass ($M^*_{sph}$), central velocity dispersion ($\sigma_{0_{sph}}$), effective radius of the spheroid ($R_{e_{sph}}$), and the density at the SMBH sphere of influence $\rho_{soi}$ measured by \cite{2022ApJ...927...67S}.
Fig.~\ref{galaxy} summarizes the definitions underlying these variables.
These variables are sufficient to characterize the SMBH and its environment, the central spheroid of the host galaxy.
Masses are measured in Solar units, lengths in parsec, and velocities in $km/s$.
Following standard practice, we consider the base-10 logarithms of these quantities.
A summary table of the sample is reported in Tab.~\ref{tab:DataSet}.
A pair plot for these variables is shown in Fig.~\ref{DataSet}.

\begin{table}[ht]
\centering
\begin{tabular}{lrrrrr}
\toprule
{} &  $\rho_{soi}$ &  $M_{BH}$ & $\sigma_{0_{sph}}$ & $R_{e_{sph}}$  &  $M^*_{sph}$ \\
\midrule
mean  &         2.70 &         8.54 &             2.33 &          0.30 &           10.84 \\
std   &         0.83 &         0.91 &             0.18 &          0.67 &            0.80 \\
min   &         0.71 &         5.74 &             1.54 &         -1.24 &            8.03 \\
max   &         4.67 &        10.30 &             2.59 &          1.49 &           12.26 \\
\end{tabular}
\vspace{0.1in}
    \caption{Summary of the early-type galaxy data set.}
    \label{tab:DataSet}
\end{table}

\section{Methods}
\subsection{PC algorithm}
The Peter and Clark (PC) algorithm \citep[][]{spirtes2000causation} is one of the most well-known and time-tested causal discovery algorithms, used to infer causal relationships from observational data.
The algorithm seeks to identify a skeletal structure of a causal graph by iteratively testing conditional independencies in the data.
Initially, it assumes a fully-connected undirected graph among all variables.
In the first phase, edges are removed based on conditional statistical independence tests, gradually increasing the conditioning set size.
After obtaining the undirected skeleton, the algorithm enters the orientation phase, where it employs a set of rules to determine the direction of the remaining edges.
However, it is not possible in general to orient all the edges, leading to a Partially Directed Acyclic Graph (PDAG), representing a Markov equivalence class of Directed Acyclic Graphs (DAGs).
This graph captures the causal relationships that are identifiable from the observed data under certain assumptions, such as faithfulness, causal sufficiency, and acyclicity.

We use the PC algorithm in the implementation by \cite{2023arXiv230716405Z} through the \emph{causal-learn} library in Python, relying on the Kernel Conditional Independence test \citep[KCI; ][]{2012arXiv1202.3775Z}. The KCI test is a non-parametric method used to assess the conditional independence between two random variables given a set of conditioning variables.
The principle underlying KCI is to exploit the properties of reproducing kernel Hilbert spaces (RKHS).
The idea is to embed the distributions of the variables into high-dimensional Hilbert spaces and then measure the independence in these spaces using a kernel-based distance metric.
Essentially, if two variables are conditionally independent given a third variable, their joint distribution can be factorized into two separate distributions, and this factorization manifests as orthogonality in the RKHS.
The KCI test statistically evaluates this orthogonality to decide whether conditional independence holds.
By using kernel methods, the KCI test can capture nonlinear dependencies.

We set the threshold for rejecting the null hypothesis of conditional independence to a p-value of $0.1$.
This is a relatively generous threshold, erring towards rejection, meaning a graph with more edges.
We also considered a value of $0.01$, obtaining slightly different results, discussed in the following. 





\section{Results}
\begin{figure}
  \centering
  \includegraphics[width=\textwidth]{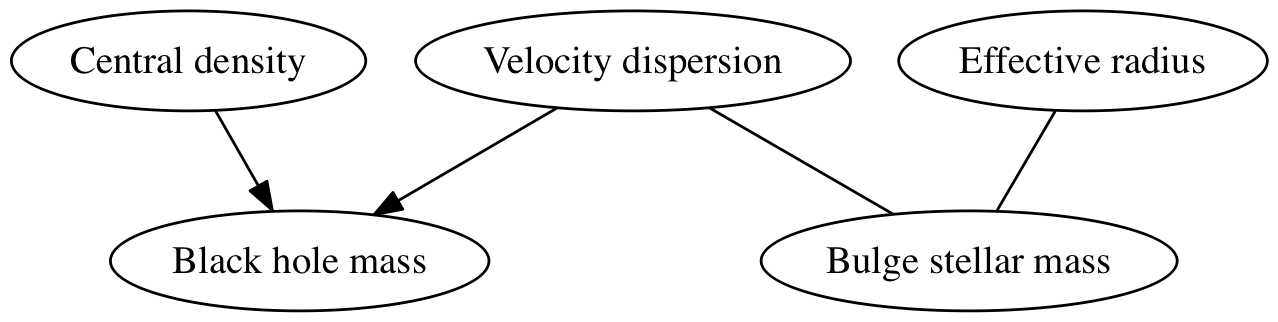}
  \caption{CPDAG learned by the PC algorithm on our data set.
  Directed edges represent causal dependencies, undirected edges represent lack of independence.\label{PC}}
\end{figure}

Fig.~\ref{PC} shows the Complete Partially Directed Acyclic Graph (CPDAG) which represents the Markov equivalence class learned by the PC algorithm on our data set.
The main result is that SMBH mass is a direct effect of central density within the sphere of influence and of velocity dispersion.


\subsection{Physical interpretation}
Bondi–Hoyle–Lyttleton accretion \citep[][]{hoyle1941accretion} on a black hole predicts that mass growth is determined by the ambient density and the velocity dispersion of the medium. Similarly, repeated mergers of seed black holes, are likely governed by a similar dependence. with encounter cross-section being determined by relative velocity in the gravitational focusing limit. This is an indication that our causal discovery approach uncovered a genuine physical relation between bulge properties and SMBH mass, especially if we admit that currently observed quantities such as the stellar velocity dispersion are likely to trace primordial quantities.


\subsection{Robustness}
The PC algorithm relies on (conditional) independence tests between variables.
With finite data, the test may fail to find a dependence that is actually present, or vice versa deem that two variables are dependent when they in fact are not.
To test this, we change the value of the threshold for conditional independence used by the KCI test from $0.1$ to $0.01$.
The resulting CPDAG is almost identical to the one found initially (shown in Fig.~\ref{PC}) with the only difference being that the edge between velocity dispersion and bulge stellar mass is missing.
The main result regarding SMBH mass being caused by central density and velocity dispersion stands unaffected.

We also test the reliability of the results we obtained by re-running our analysis on a subset of the data.
We repeat $10$ times a procedure where we randomly exclude $10\%$ of our data set and run the PC algorithm on the remaining $90\%$.
This results in five unique different CPDAGs. 
Three are essentially minor variations on the original computed on the whole data set and are presented in Fig.~\ref{goodCPDAGs}.
These account for $60\%$ of the CPDAGs we computed in this fashion. The other two are shown in Fig.~\ref{badCPDAG0} (accounting for $30\%$ of the runs) and in Fig.~\ref{badCPDAG1} (accounting for $10\%$ of the runs).
The last two causal structures differ from the one we found using the whole data set in that black hole mass affects bulge stellar mass, but the result that velocity dispersion and central density cause SMBH mass, still holds. 

\subsection{Limitations}
Early galactic evolution is rife with quantities we are not able to measure directly. The causal sufficiency assumption underlying the PC algorithm, that is the absence of unobserved confounds, is, therefore, a strong assumption to make. There is some precedent in the literature suggesting that, indeed, observed correlations may be the result of a common mechanism of accretion, rather than a direct causal link \citep[][]{2011ApJ...734...92J}. An additional source of confounds is the dynamical evolution of the system itself: past values of any variable may in principle causally affect both the current values of that variable and the current values of other variables.

\begin{figure}
  \centering
  \includegraphics[width=\textwidth]{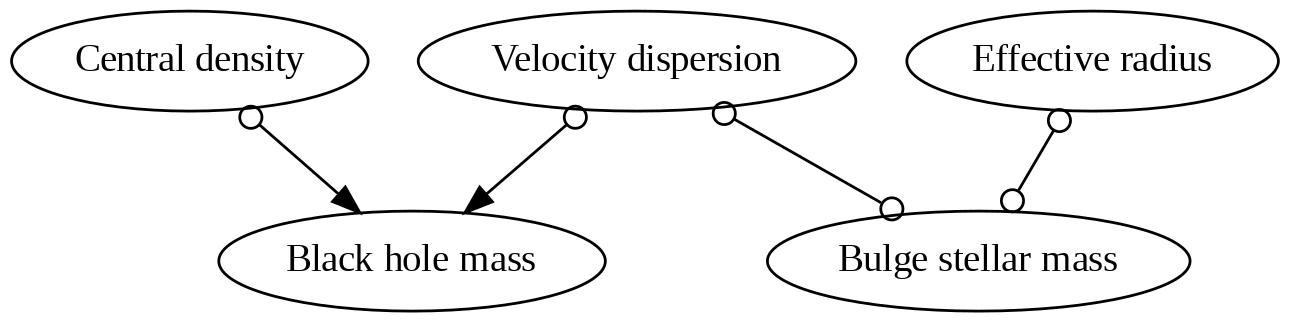}
  \caption{PAG learned by the FCI algorithm on our data set. Directed edges represent causal dependencies, undirected edges represent lack of independence, like in Fig.~\ref{PC}. However here empty circles denote a relation that may be due to an unobserved confound. \label{FCI}}
\end{figure}

To address the issue of unobserved confounds we applied the Fast Causal Discovery \citep[FCI; ][]{spirtes2000causation} algorithm to our data set. FCI learns a Partial Ancestral Graph (PAG), shown for our data in Fig.~\ref{FCI}. PAGs contain additional edge types with respect to CPDAGs. In particular the edges marked with an empty circle represent a situation where there is an association between variables (the null hypothesis of statistical independence has been rejected) but the nature of the association may not be causal. If this is the case, the association is due to one or more unobserved confounds. The FCI algorithm on our dataset indeed cannot decide whether central density and velocity dispersion cause SMBH mass (as found by the PC algorithm) or whether the observed association is due to unobserved confounds. It does however rule out SMBH mass causing central density and velocity dispersion.

\section{Conclusions}
We applied causal discovery via the PC and FCI algorithms for the first time to a state-of-the-art data set of dynamically measured SMBHs and their host early-type galaxy properties.
The causal structure we learned from this data suggests that SMBH mass is the effect of central velocity dispersion and density at the SMBH sphere of influence.
The result is robust to changes in hyperparameters such as the threshold for statistical significance in the KCI independence test used by the PC algorithm and to random subsampling of our data set.
From the physical point of view, this suggests that the bulge properties determine the SMBH mass by controlling SMBH assembly, which likely takes place through repeated mergers of seed black holes.

\begin{ack}
M. P. acknowledges financial support from the European Union’s Horizon 2020 research and innovation program under the Marie Skłodowska-Curie grant agreement No. 896248. This material is based upon work supported by Tamkeen under the NYU Abu Dhabi Research Institute grant CASS.

\end{ack}


\bibliography{neurips.bib}
\bibliographystyle{plain}

\appendix
\section{Additional figures}
\begin{figure}[ht]
  \centering
  \begin{tabular}{c}
  \includegraphics[width=0.8\textwidth]{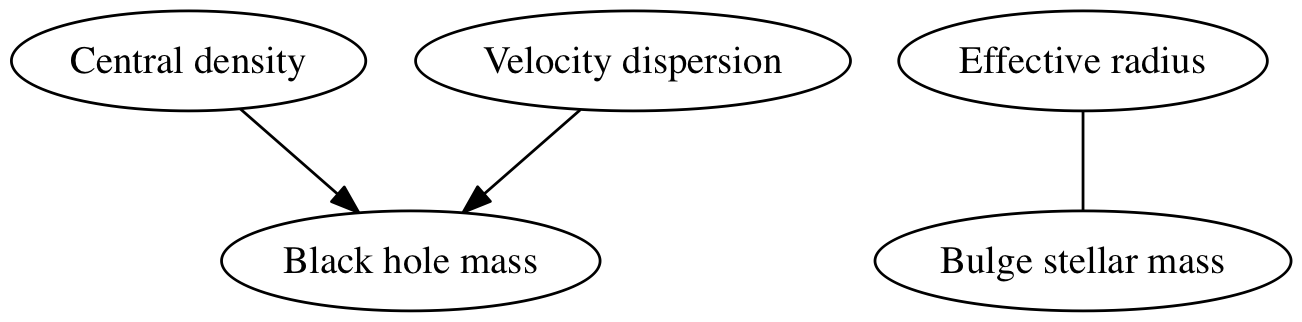} \\
  \vspace{0.2in}\\
  \includegraphics[width=0.8\textwidth]{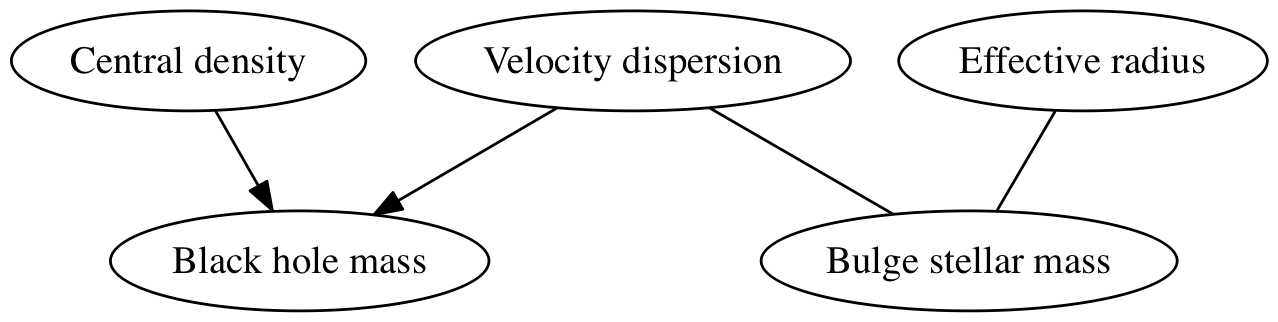} \\
  \vspace{0.2in}\\
  \includegraphics[width=0.8\textwidth]{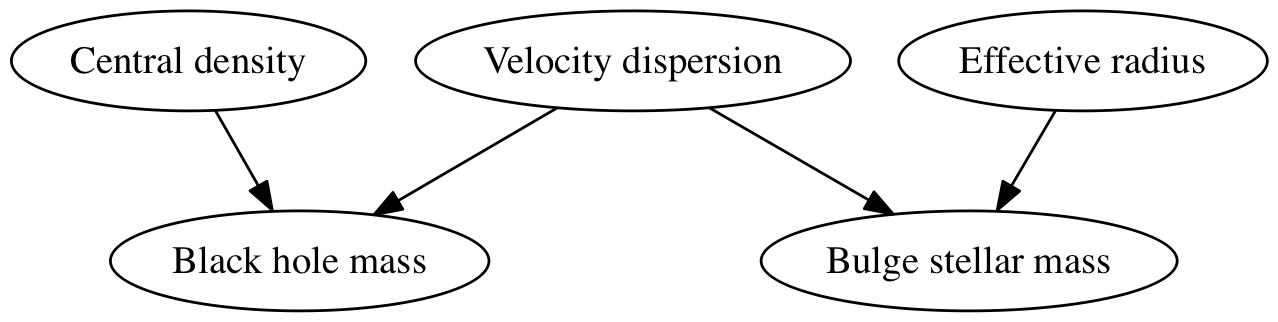} \\
  \vspace{0.2in}\\
  \end{tabular}
  \caption{Three CPDAGs learned by the PC algorithm on a random subsample containing $90\%$ of the original data set.\label{goodCPDAGs}}
\end{figure}

\begin{figure}[ht]
  \centering
  \includegraphics[width=0.8\textwidth]{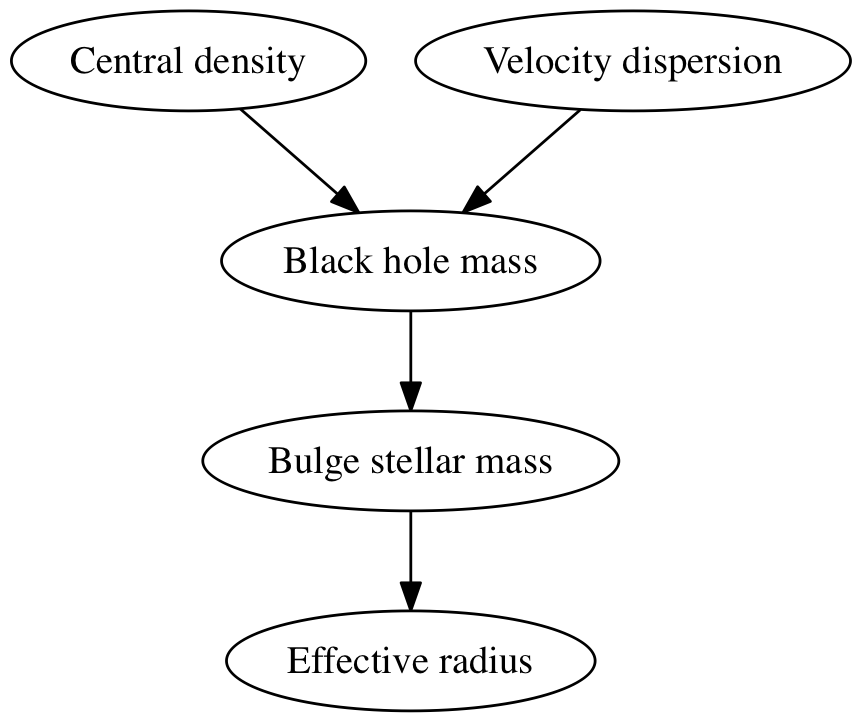} \\
  \caption{CPDAG learned by the PC algorithm on a random subsample containing $90\%$ of the original data set.\label{badCPDAG0}}
\end{figure}

\begin{figure}[ht]
  \centering
  \includegraphics[width=0.8\textwidth]{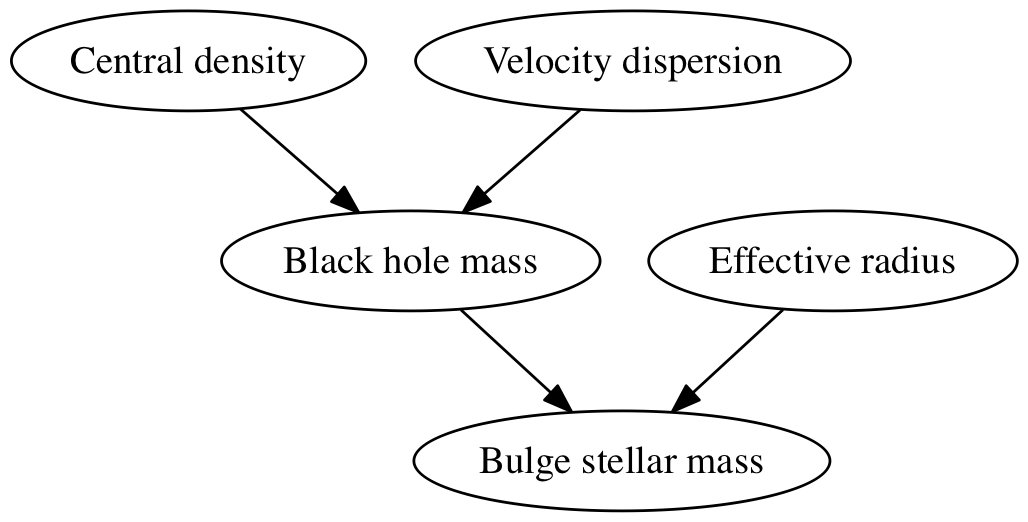} \\
  \caption{CPDAG learned by the PC algorithm on a random subsample containing $90\%$ of the original data set.\label{badCPDAG1}}
\end{figure}

\end{document}